\documentclass[useAMS,usenatbib]{mn2e}
\usepackage{graphicx}
\usepackage{amssymb}
\input{journaldef.sty}

\newcommand{\nupeak}{$\nu_{peak}$} 
\newcommand{\speak}{$S_{peak}$} 
\newcommand{\lpeak}{$L_{peak}$} 

\title[CSS and GPSs: Synchrotron Self Absorption]
{Numerical calculations of spectral turnover and Synchrotron Self Absorption in CSS and GPS radio sources.}

\author[S. Jeyakumar]{S. Jeyakumar\thanks{E-mail:
sjkastro.ugto.mx} \\
Departamento de Astronom{\'i}a, Universidad de Guanajuato, AP 144, Guanajuato CP 36000, M{\'e}xico. \\
}

\begin{document}
\setcounter{table}{0}

\date{}

\pagerange{\pageref{firstpage}--\pageref{lastpage}} \pubyear{}

\maketitle

\label{firstpage}

\begin{abstract}
The dependence of the turnover frequency
on the linear size is presented  for a sample of GPS and CSS 
radio sources derived from complete samples. 
The dependence of the luminosity of the emission at 
the peak frequency with the linear size and the peak frequency 
is also presented for the galaxies in the sample. 
The luminosity of the smaller sources evolve strongly with 
the linear size.  Optical depth effects have been 
included to the 3D model for the radio source of Kaiser (2000)
to study the spectral turnover. Using this model, the observed 
trend can be explained by synchrotron self absorption. 
The observed trend in the peak-frequency -- linear-size plane is 
not affected by the luminosity evolution of the sources. 
\end{abstract}


\section{Introduction}
The current consensus is that the 
Compact Steep Spectrum (CSS) and  Giga-hertz Peaked Spectrum (GPS)
sources are young radio sources,  evolving in the 
Inter-Stellar Medium (ISM) of the host galaxy \citep[cf.][]{fanti.etal95, carvalho98, odea98,
murgia.etal99, sjk.etal05}. 
These classes of radio sources provide the opportunity to study
the evolution of the radio sources
in their youth and the interaction of the radio jets and lobes with the
material in the ISM. The GPSs are selected based on the turnover in the spectrum.
Although the CSSs are not selected based on such a turnover,
the spectra of most of the CSSs show either a turnover or flattening 
at very low frequencies \citep{fanti.etal90}. 
The processes that can produce turnover in the spectrum are 
synchrotron self-absorption (SSA),  free-free absorption (FFA) and
Induced Compton Scattering (ICS) \citep{odea&baum97,bicknell.etal97, kuncic.etal98}.
Studies of a few individual sources report evidence of FFA in them
\citep{bicknell.etal97, xie.etal05, kameno.etal03a, kameno.etal03b}.
\citet{muto02}  find that the polarisation properties are not 
according to the trend expected from SSA.  However using a sample of 
GPS galaxies \citet{snellen.etal00} find that the turnover is 
consistent with SSA. 

The intrinsic turnover frequency (\nupeak)  is found to be anti-correlated 
with the linear size (LS) of the source, in the complete samples of CSS
and GPS sources (O'Dea \& Baum 1997, hereafter OB97; Fanti et al. 1990).
It has been shown that FFA by the material ionised by the fast moving cocoon 
of the young radio source can explain this trend \citep{bicknell.etal97}. 
Alternatively, using a model for the evolution of the source  
\citet{odea&baum97}  could predict such an anti-correlation using
a simple homogeneous SSA model. However the observed trend is 
flatter than that predicted by SSA. This could be due 
to the evolution of the luminosity of the radio source resulting
in the observed trend \citep{odea&baum97} or due to the picture 
of SSA by a homogeneous medium being too simple. 
So, in this work a 3D model is used to study the effects of the evolution 
of the radio source on the observed trend in the \nupeak\ --~LS plane, 
in the context of SSA. 
New samples of GPS  and CSS sources have become 
available in recent years which have been  used to  extend 
the \nupeak\ --~LS plane. In addition, the luminosity evolution of the 
sources is also studied.
The sample of sources is described in Section~\ref{sec:sample}. 
The SSA model and the model results are described in 
Sections~\ref{sec:model} and \ref{sec:results} respectively. 
A summary of the results is presented in Section~\ref{sec:conclusions}.

\section{The sample of sources}
\label{sec:sample}
The sample of sources is derived from the complete samples
of CSS and GPS sources available in the literature. 
The GPS sources have well defined turnover
in the spectrum by selection. The CSS sources are selected based
on the spectral index and is not necessary to have a turnover in the spectrum. 
Since the aim here is to study the turnover in the spectrum, 
only those CSS sources with a clear turnover or flattening of
the spectrum at low frequencies are considered here. 
In the case of flattening of the spectrum at low frequencies,
the measurement at the lowest available frequency is chosen as
the turnover frequency.  Those CSSs, mostly from the B3-VLA CSS sample,
which show a curvature at low frequencies without a clear flattening 
are also not considered here \citep{fanti.etal01}. 
Variability studies of the GPS sources  have shown that most of the
quasars identified as GPSs may not be genuine peakers, but
flaring blazars \citep{tinti.etal05, tinti&zotti05, tornikoski.etal01}. 
To avoid such sources  affecting the results, 
only those GPS catalogues for which structural information is available
for most of the sources are considered. 
However galaxies from the GPS catalogues are considered 
to study the dependence of the \nupeak\  with \lpeak\  (luminosity at the 
peak frequency) even if the structural information  is not available 
\citep[cf.][]{tinti.etal05}. The samples of CSS and GPS sources used here 
are described below.

\noindent
(C) The \citet{fanti.etal90} sample of 
CSSs from the 3C catalogue. The sources in this sample which do not follow 
the above criteria are, 3C43, 3C186, 3C190, 
3C303.1, 3C305.1 and 3C455. These sources have  been removed from 
the current sample.

\noindent
(S) The \citet{stanghellini.etal98} sample of GPSs. 
These two samples are well studied. The sources from these samples
are listed in \citet{odea&baum97}.

\noindent
(F) The B3-VLA sample of CSS sources \citep{fanti.etal01}. 
Information on the structure of the sources is available for most of 
the sources (Dallacasa et al. 2002a,b). 
\nocite{dallacasa.etal02a, dallacasa.etal02b}

\noindent
(W) The sample of GPS galaxies from the WENSS catalogue \citep{snellen.etal00}.

\noindent
(A) The sample of CSS and GPS sources in the southern sky 
    \citep{edwards&tingay04}. From this sample only those galaxies 
    listed in \citet{tinti&zotti05} are considered.

\noindent
(P) The sample of confirmed CSOs \citep{peck&taylor00, gugliucci.etal05} 
are also used for this study irrespective of the liberal selection 
criteria on the spectral index,
since the CSOs are known to be young radio sources 
\citep{readhead.etal96a,owsianik&conway98, gugliucci.etal05}.  
In this sample, some of the sources show flat spectra (see below) 
making it difficult to estimate the turnover frequencies. 
Such sources have been removed from the current sample.

\noindent
(D) The galaxies from the sample of High Frequency Peakers 
    \citep{dallacasa.etal00} listed in \citet{tinti&zotti05}.

\noindent
(H) The sample of GPS galaxies from the Parkes half-Jy catalogue 
    \citep{snellen.etal02}. 

\noindent
(B) The galaxies from the \citet{bolton.etal04} sample of GPS sources listed in
\citet{tinti&zotti05}.

\noindent
(L) The GPSs from the CORALZ sample \citep{snellen.etal04}.

For the sources in the samples (C), (P) and (F) the flux density 
measurements available in the literature (from NED and the CATS database)
have been used to fit the spectrum as described in \citet{steppe.etal95}.
The values of \nupeak\  and \speak\  (flux density at the peak frequency) 
estimated from this fit are listed
in the tables. 

\begin{table}
 \begin{minipage}{8.5cm}
 \caption{The sample of GPS galaxies}
 \label{table:samplegps}
  \begin{tabular}{l c  l l l c }
   \hline
    Source & Sample$^{a}$ & z$^{b}$ & \nupeak$^c$\  & \speak\  & log(\lpeak)  \\
           &        &   &  MHz   &  Jy    &  W/Hz/Sr  \\
   \hline
       J0020+3152 & B &       1.1* &          4910 &        0.0438 &  25.29 \\
   J0032+2758 & B &      0.51* &          4600 &        0.0344 &  24.33 \\
   J0936+3207 & B &      0.20* &         14000 &         0.056 &  23.62 \\
   J1506+4239 & B &      0.38* &         14000 &         0.766 &  25.38 \\
   J1517+3936 & B &      0.46* &         21000 &         0.043 &  24.32 \\
   J1526+4201 & B &      0.36* &          8100 &         0.067 &  24.27 \\
   J1530+3758 & B &      0.19* &          3430 &         0.141 &  23.97 \\
   J1540+4138 & B &      0.17* &          8800 &         0.046 &  23.38 \\
   J1550+4536 & B &      0.50* &          3500 &         0.062 &  24.57 \\
   J1554+4350 & B &       1.2* &         10900 &         0.045 &  25.40 \\
 J0108$-$1201 & H &       1.0* &          1000 &           0.9 &  26.49 \\
 J0206$-$3024 & H &      0.65* &           500 &           0.9 &  26.01 \\
   J0210+0419 & H &       1.5* &           400 &           1.3 &  27.14 \\
 J0210$-$2213 & H &       1.4* &          1500 &           1.1 &  26.98 \\
 J0242$-$2132 & H &     0.314  &          1000 &           1.3 &  25.42 \\
   J0323+0534 & H &      0.37* &           400 &           7.1 &  26.32 \\
 J0401$-$2921 & H &      0.65* &           400 &             1 &  26.05 \\
 J0407$-$3924 & H &      0.54* &           400 &           1.4 &  26.00 \\
 J0407$-$2757 & H &      0.68* &          1500 &           1.4 &  26.25 \\
 J0433$-$0229 & H &      0.36* &           400 &             3 &  25.92 \\
 J0441$-$3340 & H &      0.65* &          1500 &           1.2 &  26.13 \\
 J0457$-$0848 & H &      0.52* &           400 &             1 &  25.82 \\
   J0913+1454 & H &      0.47* &           600 &           1.1 &  25.75 \\
 J1044$-$2712 & H &      0.65* &          1500 &           0.8 &  25.96 \\
   J1057+0012 & H &      0.65* &           400 &           1.6 &  26.26 \\
   J1109+1043 & H &      0.55* &           500 &           2.4 &  26.25 \\
 J1110$-$1858 & H &     0.497  &          1000 &           0.9 &  25.72 \\
 J1122$-$2742 & H &      0.65* &          1400 &           0.8 &  25.96 \\
 J1135$-$0021 & H &      0.16* &           400 &           2.9 &  25.12 \\
   J1203+0414 & H &      0.33* &           400 &           1.4 &  25.50 \\
 J1345$-$3015 & H &      0.65* &           400 &           2.5 &  26.45 \\
 J1350$-$2204 & H &      0.63* &           400 &           1.4 &  26.17 \\
   J1352+0232 & H &      0.47* &           400 &             2 &  26.01 \\
   J1352+1107 & H &      0.65* &           400 &           3.6 &  26.61 \\
 J1447$-$3409 & H &      0.65* &           500 &             1 &  26.05 \\
 J1506$-$0919 & H &      0.43* &           600 &           1.6 &  25.82 \\
 J1548$-$1213 & H &     0.883  &           400 &           3.7 &  26.96 \\
 J1556$-$0622 & H &      0.94* &           400 &           2.4 &  26.84 \\
 J1600$-$0037 & H &            &          1000 &           1.2 &  27.10 \\
 J1604$-$2223 & H &     0.141  &           600 &             1 &  24.54 \\
   J1640+1220 & H &     1.150  &           400 &           3.7 &  27.27 \\
   J1648+0242 & H &      0.65* &           400 &           3.4 &  26.58 \\
   J2058+0540 & H &     1.381  &           400 &           3.1 &  27.41 \\
 J2123$-$0112 & H &     1.158  &           400 &             2 &  27.01 \\
   J2151+0552 & H &     0.740  &          5000 &           1.2 &  26.27 \\
 J2325$-$0344 & H &       1.4* &          1400 &           1.2 &  27.02 \\
 J2339$-$0604 & H &       1.2* &           400 &           3.8 &  27.33 \\
   J0733+5605 & L &     0.104  &           460 &          0.42 &  23.88 \\
   J0739+4954 & L &     0.054  &           950 &           0.1 &  22.67 \\
   J0831+4608 & L &     0.127  &          2200 &          0.13 &  23.56 \\
   J0906+4636 & L &     0.085  &           680 &           0.3 &  23.56 \\
   J1317+4115 & L &     0.066  &          2300 &          0.27 &  23.28 \\
   J1718+5441 & L &     0.147  &           480 &          0.44 &  24.22 \\

   \hline
  \end{tabular}

\medskip

$^{a}$~See text for the meaning of the labels in the sample column. 
$^{b}$~The * denotes a photometric redshift. 
$^{c}$~Here \nupeak\ refers to the observed value\\
 \end{minipage}
\end{table}

Of these samples (H), (B) and (L) have been used to study 
only the dependence of \nupeak\  with \lpeak, since 
information on the source structure is not available for the majority of the 
sources. The sources from these three samples are listed in 
Table~\ref{table:samplegps}. The columns are arranged as follows:
column 1: source name; column 2: master sample; column 3: redshift ;
column 4: observed \nupeak\  in MHz ; column 5: \speak\  in Jy; column 6: 
log of \lpeak\  in W/Hz/Sr. 
The sources from all other samples are listed in Table~\ref{table:samplecss}. 
The columns are arranged as follows:
column 1: source name; column 2: master sample; column 3: optical id; 
column 4: redshift ; column 5: largest angular size (LAS) in arc seconds; 
column 6: observed \nupeak\  in MHz ; column 7: \speak\  in Jy; column 8: 
log of \lpeak\  in W/Hz/Sr ; column 9: references to LAS, \nupeak\  and \speak.
Sources appearing on many samples are counted only once.
A redshift of 1.5 is assumed for those sources without any redshift 
measurement.  The cosmological constants used here are, $H_0=75$~kms$^{-1}$Mpc$^{-1}$ and $q_0$=0.0. 

Other samples of CSS and GPS sources available in the literature
are, the CSS sources from the FIRST catalogue \citep{kunert.etal02}, 
the weak CSSs \citep{tschager.etal03}, the B2 sample of CSSs 
\citep{saikia.etal02},  the sample of CSSs 
from the S4 sample \citep{saikia.etal01} and the sample of GPSs from
the JVAS sample \citep{marecki.etal99}.
Since the optical identification or structural information 
is less complete in these catalogues, they are not considered here. 

The combined sample consists of a total of 203 sources of which 
150 are CSS/GPS sources and 53 are GPS galaxies. The sample size used 
in the present study is about three times larger than OB97 sample of sources.  
The upper ranges of redshift, linear size and peak luminosity 
are similar in both the samples. There are only three sources with large rest frame 
peak frequency in the current sample. The lowest redshift in this sample is 0.004 as 
compared to 0.08 in the OB97 sample.  However the present sample extends the range of 
linear sizes by an order of magnitude lower to about 0.65~pc and the peak luminosity 
by 2.5 orders of magnitude lower than the OB97 sample.

\begin{table*}
 \begin{minipage}{14cm}
 \caption{The sample of CSS \& GPS sources$^\dagger$}
 \label{table:samplecss}
  \begin{tabular}{l c l l l l l l c l}
   \hline
    Source & Sample$^{a}$ & ID & z$^{b}$ & LAS & \nupeak$^{c}$\  & \speak\  & LS & log(\lpeak)  & Ref$^{d}$\\
           &     &    &   &  '' &   MHz   &  Jy    & kpc & W/Hz/Sr & \\
   \hline
       J0022+0014 & SH &  G &     0.305  &  0.06 &           700 &          3.47 &           0.24 &  25.82 & 1,20\\
   J0111+3906 & SP &  G &     0.669  &  0.006 &          4000 &          1.33 &         0.0373 &  26.21 & 1,20\\
   J0137+3309 & C &  Q &     0.367  &  0.5 &            80 &          68.6 &           2.25 &  27.29 & 1\\
   J0141+1353 & C &  G &     0.621  &  0.92 &           100 &          10.8 &           5.53 &  27.04 & 1\\
   J0224+2750 & C &  G &     0.309  &  2.5 &            26 &            21 &           10.1 &  26.61 & 1\\
   J0226+3421 & C &  Q &     2.910  &  1.1 &           150 &           4.4 &           9.97 &  28.53 & 1,20\\
 J0240$-$2309 & S &  Q &     2.223  &  0.018 &          1000 &          7.05 &          0.158 &  28.37 & 1\\
   J0251+4315 & S &  Q &     1.316  &  0.06 &          5000 &          1.27 &          0.473 &  26.96 & 1,20\\
   J0318+1628 & SC &  G &            &  0.3 &           800 &          9.55 &           2.44 &  28.00 & 1,20\\
   J0321+1221 & C &  Q &     2.662  &  0.02 &           400 &           2.4 &          0.179 &  28.15 & 1\\
   J0348+3353 & C &  G &     0.244  &  0.25 &            40 &            21 &          0.858 &  26.38 & 1\\
   J0410+7656 & CP &  G &    0.5985  &  0.15 &           350 &           6.9 &          0.885 &  26.80 & 1\\
   J0431+2037 & SC &  G &     0.219  &  0.25 &          1100 &          4.02 &          0.793 &  25.56 & 1,20\\
   J0432+4138 & C &  G &     1.023  &  0.08 &           150 &            18 &          0.586 &  27.82 & 1\\
   J0459+0229 & S &  Q &     2.384  &  0.012 &          2100 &          1.89 &          0.106 &  27.90 & 1,16,20\\
   J0503+0203 & SH &  Q &     0.583  &  0.015 &          1800 &          2.51 &         0.0874 &  26.34 & 1,20\\
   J0521+1638 & C &  Q &     0.760  &  0.60 &           200 &          18.7 &           3.94 &  27.50 & 1\\
   J0542+4951 & C &  Q &     0.545  &  0.55 &           100 &          69.4 &            3.1 &  27.71 & 1\\
   J0713+4349 & SP &  G &     0.518  &  0.025 &          1900 &          2.09 &          0.137 &  26.13 & 1,20\\
   J0741+3112 & S &  Q &     0.630  &  0.010 &          5300 &          3.82 &         0.0605 &  26.60 & 1,20\\
   J0745+1011 & S &  G &     2.624  &  0.010 &          2700 &          4.12 &         0.0896 &  28.36 & 1,20\\
 J0745$-$0044 & S &  Q &     0.994  &  0.005 &          5800 &          2.12 &         0.0363 &  26.85 & 1,20\\
 J0943$-$0819 & SH &  G &     0.228  &  0.05 &           500 &           3.4 &          0.163 &  25.52 & 1,20\\
   J1008+0730 & C &  G &     0.877  &  1.3 &            16 &          45.5 &           9.03 &  28.04 & 1\\
   J1021+2159 & C &  G &     1.617  &  0.84 &            80 &            17 &           6.96 &  28.35 & 1\\
   J1035+5628 & SP &  Q &     0.459  &  0.040 &          1300 &          1.87 &          0.206 &  25.96 & 1,20\\
   J1120+1420 & SH &  G &     0.362  &  0.08 &           500 &          3.89 &          0.358 &  26.03 & 1,20\\
 J1130$-$1449 & S &  Q &     1.187  &  0.003 &          1000 &           5.8 &          0.023 &  27.50 & 1,20\\
 J1146$-$2447 & S &  Q &     1.950  &  0.006 &          2200 &          1.69 &         0.0515 &  27.58 & 1,20\\
   J1156+3128 & C &  Q &     1.557  &  0.9 &           150 &           8.2 &           7.39 &  27.98 & 1\\
   J1206+6413 & C &  G &     0.371  &  1.36 &            90 &          15.3 &           6.17 &  26.65 & 1\\
   J1227+3635 & C &  Q &     1.974  &  0.060 &           900 &           2.2 &          0.516 &  27.71 & 1\\
 J1248$-$1959 & S &  Q &     1.280  &  0.50 &           500 &          8.69 &           3.92 &  27.77 & 1,20\\
   J1252+5634 & C &  Q &     0.321  &  1.67 &            40 &            13 &           6.91 &  26.44 & 1\\
   J1326+3154 & SC &  G &     0.369  &  0.06 &           500 &          7.03 &          0.271 &  26.31 & 1,20\\
   J1330+2509 & C &  Q &     1.055  &  0.048 &            40 &            28 &          0.355 &  28.04 & 1\\
   J1331+3030 & C &  Q &     0.849  &  3.2 &            80 &            35 &             22 &  27.89 & 1\\
   J1347+1217 & SH &  G &     0.122  &  0.080 &           400 &          8.86 &           0.16 &  25.35 & 1,20\\
   J1400+6210 & SCP &  G &     0.429  &  0.07 &           500 &          6.56 &          0.346 &  26.43 & 1,20\\
   J1407+2827 & SD &  G &     0.077  &  0.007 &          4200 &          2.76 &        0.00936 &  24.43 & 1,20\\
   J1416+3444 & C & EF &            &  0.06 &           700 &           2.1 &          0.489 &  27.34 & 1\\
   J1419+0628 & C &  Q &     1.439  &  1.49 &            60 &            82 &             12 &  28.88 & 1\\
   J1445+0958 & SC &  Q &     3.535  &  0.02 &           900 &          2.61 &          0.184 &  28.58 & 1,20\\
   J1459+7140 & C &  Q &     0.905  &  2.11 &            40 &          40.3 &           14.8 &  28.03 & 1\\
   J1520+2016 & C &  G &     0.752  &  1.05 &            40 &            23 &           6.86 &  27.57 & 1\\
   J1521+0430 & SH &  S &     1.296  &  0.135 &           800 &          4.58 &           1.06 &  27.50 & 1,20\\
   J1602+3326 & SC &  G &     1.100  &  0.06 &          2400 &          3.06 &           0.45 &  27.13 & 1,20\\
   J1609+2641 & SC &  G &     0.473  &  0.05 &          1100 &          5.44 &          0.261 &  26.45 & 1,20\\
   J1634+6245 & C &  Q &     0.988  &  0.20 &           150 &          14.1 &           1.45 &  27.67 & 1\\
   J1638+6234 & C &  G &      0.75  &  0.24 &           150 &          14.2 &           1.57 &  27.36 & 1\\
   J1821+3942 & C &  G &       0.4  &  0.44 &           250 &           7.7 &           2.09 &  26.43 & 1\\
   J1831+2907 & C &  Q &     0.842  &  3.1 &            74 &          7.57 &           21.2 &  27.22 & 1\\
 J2011$-$0644 & SH &  G &     0.547  &  0.030 &          1400 &          2.64 &          0.169 &  26.29 & 1,20\\
 J2129$-$1538 & S &  Q &      3.27  &  0.008 &          4100 &          1.23 &         0.0733 &  28.14 & 1,20\\
   J2130+0502 & SH &  G &     0.990  &  0.030 &           700 &          4.93 &          0.217 &  27.22 & 1,20\\
   J2136+0041 & S &  Q &     1.936  &  0.002 &          4300 &          8.59 &         0.0171 &  28.28 & 1,20\\
   J2212+0152 & SH &  G &            &  0.055 &           500 &          4.51 &          0.448 &  27.68 & 1,20\\
   J2250+7129 & C &  G &     1.841  &  1.6 &            40 &            13 &           13.6 &  28.39 & 1\\
   J2251+1848 & C &  Q &     1.758  &  0.66 &            40 &            30 &           5.56 &  28.70 & 1\\
   J2344+8226 & SC &  Q &     0.735  &  0.18 &           500 &          6.29 &           1.17 &  26.99 & 1,20\\
   J2355+4950 & SP &  G &     0.237  &  0.050 &           700 &          2.93 &          0.168 &  25.50 & 1,20\\

   \hline
  \end{tabular}
 \end{minipage}
\end{table*}

\begin{table*}
 \setcounter{table}{1}
 \begin{minipage}{14cm}
  \caption{continued }
  \begin{tabular}{l c l l l l l l c l}
   \hline
    Source & Sample$^{a}$ & ID & z$^{b}$ & LAS & \nupeak$^{c}$\  & \speak\  & LS & log(\lpeak)  & Ref$^{d}$\\
           &     &    &   &  '' &   MHz   &  Jy    & kpc & W/Hz/Sr & \\
   \hline
       J0003+4807 & P & EF &            &  0.011 &          2123 &         0.348 &         0.0896 &  26.56 & 4\\
   J0132+5620 & P & EF &            &  0.012 &          3437 &           0.6 &         0.0977 &  26.80 & 4\\
   J0204+0903 & P & EF &            &  0.029 &          1061 &          1.96 &          0.236 &  27.31 & 4\\
   J0427+4133 & P & EF &            &  0.0067 &          3416 &         0.823 &         0.0546 &  26.94 & 5\\
   J0620+2102 & P & EF &            &  0.027 &          1308 &          0.86 &           0.22 &  26.96 & 4\\
   J0650+6001 & P &  Q &     0.455  &  0.00642 &          4977 &          0.88 &         0.0328 &  25.62 & 6\\
   J0753+4231 & P &  Q &      3.59  &  0.00892 &           952 &         0.727 &         0.0824 &  28.05 & 7\\
   J0754+5324 & P & EF &            &  0.0203 &          1240 &         0.634 &          0.165 &  26.82 & 4\\
   J1111+1955 & P &  G &     0.299  &  0.018 &          1305 &           1.1 &         0.0711 &  25.30 & 4\\
   J1148+5924 & P &  G &     0.011  &  0.0205 &          6149 &         0.573 &         0.0043 &  22.03 & 8\\
   J1357+4353 & P &  G &            &    &          2146 &         0.551 &    &  26.76 & 4\\
   J1414+4554 & P &  G &      0.19  &  0.030 &           693 &         0.396 &         0.0855 &  24.42 & 4\\
   J1546+0026 & PH &  G &      0.55  &  0.0083 &           365 &          2.28 &          0.047 &  26.23 & 4\\
   J1734+0926 & PH &  G &      0.61* &  0.015 &          1622 &          1.05 &         0.0893 &  26.01 & 4,16\\
   J1815+6127 & P &  Q &     0.601  &  0.0101 &           645 &         0.824 &         0.0597 &  25.88 & 7\\
   J1816+3457 & P &  G &     0.245  &  0.037 &           440 &         0.983 &          0.127 &  25.05 & 4\\
   J1823+7938 & P &  G &     0.224  &  0.0156 &          7039 &         0.554 &         0.0503 &  24.72 & 7\\
   J1845+3541 & P &  G &     0.764  &  0.0208 &          2545 &          1.07 &          0.137 &  26.26 & 17\\
   J1944+5448 & P &  G &     0.263  &  0.041 &           778 &          1.77 &          0.148 &  25.38 & 6\\
   J2022+6136 & P &  Q &    0.2280  &  0.0077 &          4086 &          2.64 &         0.0251 &  25.41 & 18\\
   J2203+1007 & P & EF &            &  0.010 &          4427 &         0.306 &         0.0814 &  26.51 & 4\\
   J0000+4054 & FP &  E &            &  0.12 &           323 &          2.06 &          0.977 &  27.33 & 3,21\\
   J0042+3739 & F &  G &     1.006  &  0.12 &           333 &           2.4 &          0.874 &  26.92 & 3,21\\
   J0042+4009 & F &  E &            &  3.7 &            74 &          4.04 &           30.1 &  27.63 & 3\\
   J0225+4229 & F &  G &       3.5* &  3.4 &           151 &          1.83 &           31.3 &  28.41 & 3\\
   J0704+3911 & F &  Q &     1.238  &  1.8 &           151 &           1.7 &             14 &  27.02 & 3\\
   J0706+4647 & F &  E &            &  0.075 &           777 &          1.81 &          0.611 &  27.28 & 3,21\\
   J0725+3917 & F &  E &            &  0.30 &           280 &          3.26 &           2.44 &  27.53 & 3,22\\
   J0758+3929 & F &  G &     2.119  &  2.6 &           151 &           3.2 &           22.6 &  27.97 & 3\\
   J0804+4704 & F &  E &            &  1.0 &           151 &           3.6 &           8.14 &  27.58 & 3\\
   J0812+4019 & F &  G &     0.551  &  1.2 &           151 &          3.47 &            6.8 &  26.42 & 3\\
   J0825+3919 & F &  G &      1.18  &  0.07 &           517 &          1.77 &          0.536 &  26.98 & 3,21\\
   J0843+4215 & F &  E &            &  0.135 &           412 &          2.13 &            1.1 &  27.35 & 3,21\\
   J0958+3848 & F &  E &            &  4.8 &            74 &          2.44 &           39.1 &  27.41 & 3\\
   J1010+4159 & F &  E &            &  0.265 &           116 &           1.5 &           2.16 &  27.20 & 3,21\\
   J1011+4204 & F &  E &            &  0.115 &           424 &          1.16 &          0.937 &  27.08 & 3,21\\
   J1017+3901 & F &  G &     0.206  &  6.1 &            74 &          7.08 &           18.5 &  25.75 & 3\\
   J1019+4408 & F &  G &      0.33* &  0.155 &           342 &          1.17 &          0.653 &  25.42 & 3,21\\
   J1030+3857 & F &  E &            &  1.6 &           107 &          1.28 &             13 &  27.13 & 3\\
   J1047+4508 & F &  G &       4.1* &  1.0 &           209 &          2.32 &           9.32 &  28.74 & 3\\
   J1052+3811 & F &  G &     1.018  &  0.21 &           429 &          1.33 &           1.54 &  26.68 & 3,21\\
   J1058+4010 & F &  E &            &  2.8 &            74 &          2.78 &           22.8 &  27.47 & 3\\
   J1131+4514 & F &  G &       0.4  &  0.5 &           151 &           6.3 &           2.37 &  26.34 & 3\\
   J1135+4258 & F &  E &            &  0.045 &          1044 &          1.41 &          0.367 &  27.17 & 3,21\\
   J1139+3803 & F &  E &            &  0.07 &           335 &         0.891 &           0.57 &  26.97 & 3,21\\
   J1143+4621 & F &  G &      0.06  &  8.1 &           151 &          5.46 &           8.64 &  24.50 & 3\\
   J1200+4548 & F &  G &     0.742  &  0.62 &           258 &          2.89 &           4.03 &  26.66 & 3,22\\
   J1201+3919 & F &  G &      2.37  &  0.07 &           467 &         0.934 &          0.619 &  27.58 & 3,21\\
   J1204+3912 & F &  G &     0.445  &  2.1 &            74 &          1.94 &           10.6 &  25.94 & 3\\
   J1207+3954 & F &  G &     2.066  &  1.6 &            74 &          3.39 &           13.9 &  27.96 & 3\\
   J1214+3748 & F &  G &       1.5* &  0.47 &            74 &          3.42 &           3.83 &  27.56 & 3,22\\
   J1227+4400 & F &  G &      0.22* &  0.42 &           103 &          1.68 &           1.34 &  25.18 & 3,21\\
   J1244+4048 & FP &  Q &     0.811  &  0.07 &           405 &          2.03 &          0.472 &  26.60 & 3,21\\
   J1343+4343 & F &  E &            &  0.13 &           338 &          1.22 &           1.06 &  27.11 & 3,21\\
   J1345+3823 & F &  Q &     1.844  &  0.13 &           412 &          1.72 &            1.1 &  27.52 & 3,21\\
   J1434+4236 & F &  E &            &  0.05 &            74 &          1.67 &          0.407 &  27.24 & 3,21\\
   J1442+4044 & F &  E &            &  0.13 &           292 &          1.55 &           1.06 &  27.21 & 3,21\\
   J1451+4154 & F &  E &            &  0.12 &           296 &          2.19 &          0.977 &  27.36 & 3,21\\
   J2303+4439 & F &  G &       1.7* &  0.5 &            74 &          7.05 &           4.18 &  28.03 & 3,22\\
   J2304+4028 & F &  E &            &  0.6 &           151 &          3.74 &           4.89 &  27.59 & 3\\
   J2307+3802 & F &  G &       0.4* &  0.15 &           151 &          2.72 &          0.712 &  25.98 & 3,21\\

   \hline
  \end{tabular}
 \end{minipage}
\end{table*}

\begin{table*}
 \setcounter{table}{1}
 \begin{minipage}{14cm}
  \caption{continued }
  \begin{tabular}{l c l l l l l l c l}
   \hline
    Source & Sample$^{a}$ & ID & z$^{b}$ & LAS & \nupeak$^{c}$\  & \speak\  & LS & log(\lpeak)  & Ref$^{d}$\\
           &     &    &   &  '' &   MHz   &  Jy    & kpc & W/Hz/Sr & \\
   \hline
       J2332+4030 & F &  E &            &  0.10 &           162 &          2.13 &          0.814 &  27.35 & 3,21\\
   J0404+6050 & W &  G &            &  0.0044 &          1000 &         0.184 &         0.0358 &  26.29 & 9\\
   J0440+6157 & W &  G &            &  0.0171 &          1000 &         0.237 &          0.139 &  26.40 & 9\\
   J0541+6745 & W &  G &       1.5* &  0.0040 &          5700 &         0.192 &         0.0326 &  26.30 & 9\\
   J0544+6201 & W &  G &       1.4* &  0.0061 &          1900 &         0.129 &         0.0489 &  26.05 & 9\\
   J0834+5803 & W &  G &     0.093  & $<$0.0043 &          1600 &         0.065 &        0.00679 &  22.97 & 9\\
   J1525+6751 & W &  G &       1.1* &  0.0224 &          1800 &         0.163 &          0.168 &  25.86 & 9\\
   J1552+6813 & W &  G &       1.3* &  0.0025 &          1500 &         0.052 &         0.0197 &  25.56 & 9\\
   J1557+6211 & W &  G &       0.9* & $<$0.0044 &          2300 &         0.049 &         0.0308 &  25.10 & 9\\
   J1600+7123 & W &  G &            &  0.0227 &          1700 &         0.346 &          0.185 &  26.56 & 9\\
   J1623+6624 & W &  G &     0.201  & $<$0.0029 &          4000 &         0.363 &        0.00862 &  24.43 & 9\\
   J1655+6441 & W &  G &            &  0.0248 &          1000 &         0.069 &          0.202 &  25.86 & 9\\
   J1841+6718 & W &  G &     0.486  &  0.0061 &          2100 &         0.178 &         0.0324 &  25.00 & 9\\
   J1941+7221 & W &  G &       1.1* &  0.0315 &          1400 &         0.233 &          0.236 &  26.01 & 9\\
   J1945+7055 & WP &  G &     0.101  &  0.0319 &          1800 &         0.929 &         0.0541 &  24.20 & 9\\
 J0241$-$0815 & A &  G &     0.004  &  0.041 &          7500 &          2.71 &        0.00316 &  21.82 & 10,12\\
 J1543$-$0757 & AH &  G &     0.172  &  0.050 &           700 &          1.65 &          0.132 &  24.94 & 10\\
 J1658$-$0739 & A &  G &            &  0.007 &          4800 &          1.32 &          0.057 &  27.14 & 10\\
 J1723$-$6500 & A &  G &     0.014  &  0.007 &          2700 &          4.48 &        0.00186 &  23.13 & 10,14\\
 J1726$-$6427 & A &  G &            &    &          1100 &          3.96 &    &  27.62 & 10\\
 J1744$-$5144 & A &  G &            &  0.052 &          1000 &          6.94 &          0.424 &  27.86 & 10\\
 J1939$-$6342 & A &  G &     0.183  &  0.042 &          1400 &            15 &          0.116 &  25.96 & 10\\
 J2257$-$3627 & A &  G &     0.006  & $<$0.100 &          2700 &          1.37 &         0.0115 &  21.88 & 10,13\\
 J2336$-$5236 & A &  G &            &    &          1100 &          1.97 &    &  27.32 & 10\\
   J0428+3259 & D &  G &       0.3  &    &          7300 &         0.545 &    &  24.99 & 23\\
   J0655+4100 & D &  G &   0.02156  &  0.0016 &          7800 &          0.33 &       0.000648 &  22.38 & 23,11,19\\
   J1511+0518 & D &  G &     0.084  &  0.0049 &         11100 &         0.778 &        0.00708 &  23.96 & 23,15\\
   J1735+5049 & D &  G &            &  0.00336 &          6400 &         0.972 &         0.0274 &  27.01 & 23,6\\

   \hline
  \end{tabular}

\medskip
$^\dagger$ A machine readable version of this table is available in CDS.
$^a$~See text for the meaning of the labels in the sample column. 
$^b$~The * denotes a photometric redshift.
$^c$~ Here \nupeak\ refers to the observed value.
$^d$~The references are: (1) \citet{odea98}; 
(2) \citet{taylor&vermeulen97};
(3) \citet{fanti.etal01}; (4) \citet{peck&taylor00}; 
(5) \citet{gugliucci.etal05}; (6) \citet{xu.etal95}; 
(7) \citet{taylor.etal94}; (8) \citet{taylor.etal98}; 
(9) \citet{snellen.etal00}; (10) \citet{edwards&tingay04}; 
(11) \citet{henstock.etal95}; (12) \citet{kameno.etal01};
(13) \citet{tingay.etal03};  (14) \citet{ojha.etal04} ;
(15) \citet{xiang.etal02};  
(16) \citet{stanghellini.etal99};  
(17) \citet{polatidis.etal95};  
(18) \citet{kellermann.etal98};  
(19) \citet{dallacasa.etal00};
(20) \citet{stanghellini.etal98};  
(21) Dallacasa  et al. (2002a); \nocite{dallacasa.etal02a}
(22) Dallacasa  et al. (2002b); \nocite{dallacasa.etal02b}
(23) \citet{tinti.etal05};  
 \end{minipage}
\end{table*}

\section{Model for spectral turnover}
\label{sec:model}
It is well known that SSA can produce 
turnover in the  spectrum. The \nupeak\  occurs at a frequency 
where the SSA optical depth becomes unity.  
The SSA optical depth, $\tau_{SSA}=\alpha_\nu R$, 
where R is the path length through the source and 
$\alpha_\nu$ is the synchrotron self absorption coefficient which is 
given in Eqn.~\ref{eqn:alphanu}. 
The $\tau_{SSA}$ depends on the electron number density, the magnetic field and 
the path length, which all vary as the source evolves. 
Thus a model for the evolution of the radio source is required to 
understand the evolution of \nupeak\  with LS. 

In the recent years, semi-analytical models
of dynamical and spectral evolution of classical radio sources 
have been constructed (Kaiser \& Alexander 1997, hereafter KA;
Kaiser, Dennett-Thorpe \& Alexander 1997, hereafter KDA; Kaiser 2000, hereafter
K00; Blundell et al. 1999; Manolakou \& Kirk 2002),
based on self-similarity \citep{falle91}. 
\nocite{kaiser&alexander97, kaiser.etal97, kaiser2000,blundell.etal99,manolakou&kirk02}
These models provide a very useful tool to study the evolution of radio 
emission from the radio sources evolving in a power-law ambient medium. 
Of these models,  K00 provide a method of constructing
the 3-dimensional emissivity of the cocoon \citep[see also][]{chyzy97}.

Here, 3-dimensional  synchrotron self-absorption coefficient, 
$\alpha_\nu$, has been constructed using the approach of K00. 
This has been used to evolve the radio spectrum with the age of the source,
including synchrotron self absorption. The method employed here and 
the parameters involved are described below.

The radio source with jet power, $Q_0$ evolves in an ambient medium of
density $\rho(L_j) = \rho_0a_0^\beta L_j^{-\beta}$, where $L_j$ is the length
of the jet. The explicit expressions for the length $L_j$ is given by KA. 
The synchrotron emissivity of a small volume element
which was injected into the cocoon at time $t_i$ and has evolved to the 
present time $t$, can be written as
\begin{eqnarray}
\epsilon(\nu,t, t_i)  = { c\sigma_T \over 6\pi\nu} \gamma(t)^3 u_b(t,t_i) n(\gamma,t, t_i) 
\end{eqnarray} 
where $u_b(t,t_i)$ is the energy density of the magnetic field and 
$n(\gamma, t, t_i)$ is the electron energy spectrum.
The synchrotron self absorption coefficient is, 
\begin{eqnarray}
 \label{eqn:alphanu}
 \alpha_\nu  =  K_0 n_0 B^{(p+2)/2} \nu^{-(p+4)/2}
\end{eqnarray}
 where $K_0$ is a constant \citep{shu}. 
The scaling factor $n_0 = n(\gamma,t,t_i)/\gamma^{-p}$, where
$p$ is the power-law energy index of the electrons.
The lower and upper bound of the $\gamma$ factor  at injection are 
$\gamma_{min}$ and $\gamma_{max}$ respectively.

The equations for quantities $u_b$, $n(\gamma)$ injected at time $t_i$
and their evolution to the present time $t$ are given in KDA. 
In order to construct a 3D model, K00 uses a specialised
geometry for the cocoon and assume that the volume element injected at 
time $t_i$  is located at $l$ (in units of L$_j$) as a thin cylindrical 
volume of thickness $\delta l$ and radius $r_c(l, \alpha_1, \alpha_2)$
(equation~12 of K00). 
The location of the cylindrical volume element $l$ is given by $(t_i/t)^{\alpha_3}$.
Since all the $\delta l$ have to add up to the length of the jet, 
the constant $\alpha_3$ can be calculated self consistently 
for the assumed geometry parameters $\alpha_1$ and $\alpha_2$, 
using the equation 15 of K00. 

Using the above approach, $\epsilon_\nu$ and $\alpha_\nu$ are obtained
at the position $(l,y)$, where $y$ is the perpendicular distance
from the jet axis.  These quantities are assumed to be axi-symmetric
with respect to the jet axis. 
If the source is viewed edge-on, the path length for a ray passing 
through the position $(l,y)$ is, $R(l,y)=2\sqrt{(r_c(l)^2 - y^2}$ 
and  $r_c(l)$ is the cocoon radius at $l$. 
The surface brightness, including the optical depth effects,  
along this ray can be written as, 
\begin{eqnarray}
 S_\nu(l,y)  =  {\epsilon(\nu,l,y)\over \alpha(\nu,l,y)} \left\{ 1- exp\left[-\alpha(\nu,l,y)R(l,y)\right]\right\}
\end{eqnarray}

The total emission from the cocoon, obtained by integrating over the
surface, is given by
\begin{eqnarray}
\label{eqn:pnu}
 P_\nu  = 4 \int_{l_{low}}^1 L_j(t) dl \int_0^{r_c(l)}  S_\nu(l,y) dy
\end{eqnarray}
Here $l_{low}$ depends on whether that part of the cocoon can have 
electrons with $\gamma$ corresponding to  the given frequency, had 
that value at injection been $\gamma_{max}$.

In the model of KA, the jet length has to be large enough to allow 
for the pressure balance between the cocoon and the jet that passed through
the re-confinement shock. For a source size of 10~pc and $\beta=1.9$ 
the ratio of the location of the re-confinement shock to the
length of the jet, $R_{conf}$, is about 0.1. 
Thus the jet length is much larger than the location of the 
re-confinement shock and the model results can be applied to such sources.  
However, the density at the inner region of the host galaxy is not expected 
to vary as steeply as in the outer regions. A realistic situation
is a King type profile with a constant density core.
The models with shallower density profile $\beta \sim 0.0$ 
are a good representation for the smaller sources. 
For $R_{conf}$ of about 0.3 and $\beta$ of 0.0 and 0.5 the 
size of the source is about 150 and 50~pc respectively. 
Therefore at smaller sizes the model results are to be treated as extrapolation 
of the trend at larger sizes. 

\begin{figure}
  \includegraphics[width=8cm,clip=]{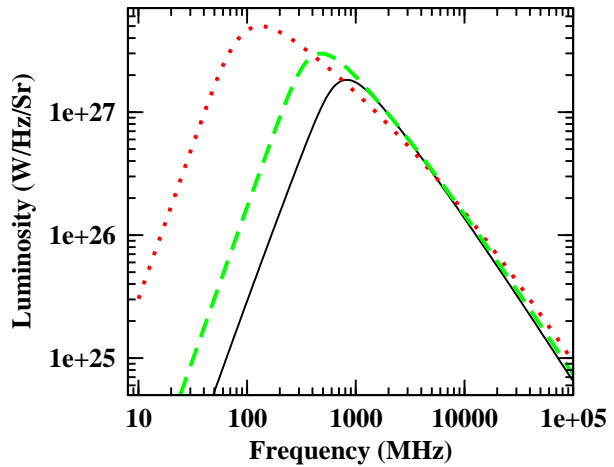}
  \caption{
The model spectra for radio sources of sizes  1 (solid),
2 (dashed) and 10 (dotted) kpc are plotted. 
The parameters used for the model calculations are 
$Q_0=0.13\times10^{40}$,  $a_0= 2$~kpc  and $\beta=1.9$.
} \label{fig:spec}
\end{figure}

\begin{figure}
  \includegraphics[width=8cm,clip=]{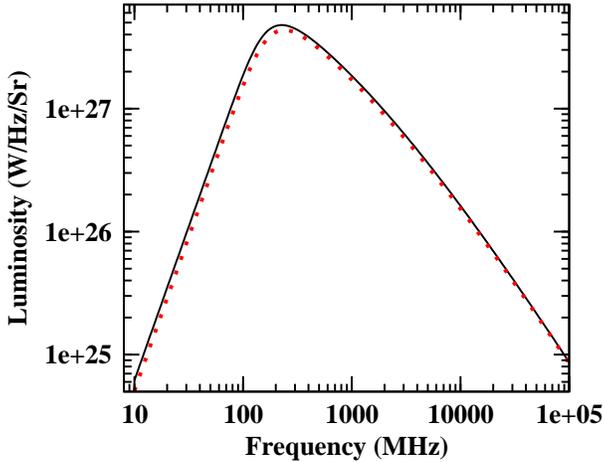}
  \caption{
The model spectra for a source with linear size of 4.88~kpc. 
The solid line represents the model 
with $\beta=0$ and the dotted line represents $\beta=1.9$.
The parameters used for the model calculations are 
as in Fig.~\ref{fig:spec}. 
} \label{fig:speccross}
\end{figure}

\citet{alexander00} has presented a method of switching from 
one power-law model to another that can apply
to evolution from smaller to larger sources and has shown that the
luminosity from different models match very close to the point of switching.
It turns out that the location where the pressure from different
models match is same as the location where the analytical 
luminosities of \citet{alexander00} match. 
So model results are also considered by switching from 
$\beta=0$ model to $\beta=\beta_{outer}$ 
at the distance where the pressure of the cocoon of different 
power-law models match. 
For $\beta_{outer}=1.9$, the pressures match at $L_j =  1.22~a_0$.
Although, all the quantities involved in 
estimating the luminosity of the source are related to the pressure, 
the continuity of the evolutionary tracks in the 
\nupeak\ --~\lpeak\  or the \nupeak\ --~LS plane is not guaranteed.  In addition,
the evolutionary history of the energy of the electrons of the two
power-law models are different. 
However it can be seen from Fig.~\ref{fig:speccross} 
that indeed the luminosity and the \nupeak\  are very close to each other at
the switching point, but not exactly the same. Results obtained using 
this approach are also presented in the next section.

\begin{figure}
  \includegraphics[width=8.5cm]{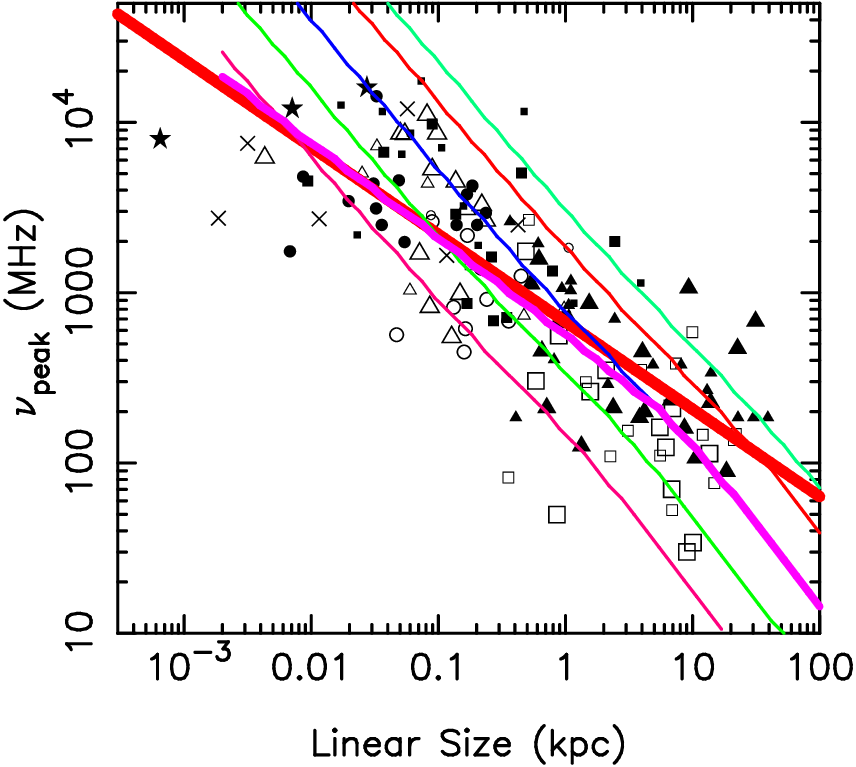}
  \caption{The \nupeak, is plotted against
the linear size of the radio source. The symbols representing the samples
are: 
unfilled square $-$ C; 
filled squre $-$ S;
unfilled triangle  $-$ P; 
filled triangle $-$ F;
unfilled circle $-$ H ; 
filled circle $-$ W;
unfilled star L; 
filled star $-$ D; 
plus $-$ B ;
cross $-$ A. 
The quasars are plotted with symbol sizes smaller 
than that of the galaxies. 
The thick solid line represents the 
least squares fit to the data. 
The thin lines from the bottom to the top of the figure represent
models with $Q_0$$=$$1.3\times10^{37}$, $1.3\times10^{38}$, 
$1.3\times10^{39}$, $1.3\times10^{40}$ and $5\times10^{50}$~W respectively. 
The model with $\beta=0$ at the inner core and $\beta=1.9$ at the
outer regions is plotted with a line of intermediate thickness. 
\label{fig:nupeakls} } 
\end{figure}

\section{Results}
\label{sec:results}
The model spectrum is obtained  by numerically integrating Eqn.~\ref{eqn:pnu}. 
The geometry parameters used in the model calculations are,
$\alpha_1=2$ and  $\alpha_2=1/3$. A very high value is chosen 
for $\gamma_{max}$ and $\gamma_{min}=1.0$. In addition, the value 
of the hotspot to the cocoon pressure is taken as $4R_T^2$, where $R_T = 2$. 
Other parameters assumed for the model calculations correspond to
the case-2 of KDA.
The model spectrum is presented in Fig.~\ref{fig:spec}, 
for a radio source of size 1,  2  and  10 kpc for $\beta=1.9$. 
The figure shows that the \nupeak\  shifts to 
low frequencies as the source grows in size. The spectral index
at the optically thick side of the spectrum is about 5/2. 
The point at which the pressure of the cocooon of the models
with $\beta=0$ and 1.9 match, corresponds to a source size of 4.88~kpc.
The model spectrum for a source with this size  is presented for 
$\beta=0$ and 1.9 in Fig.~\ref{fig:speccross}. This figure shows
that the spectra from the two different power-law models are 
indeed close to each other. 

\subsection{Dependence of \nupeak\  with  LS}

The value of \nupeak\  is plotted against the
LS in Fig.~\ref{fig:nupeakls},
for the quasars and the galaxies from the combined sample listed in 
Table~\ref{table:samplecss}. The meaning of the different
symbols are explained in the caption.
A linear fit to all the data gives a slope of  -0.51$\pm$0.03, which
is shown as thick solid line in the figure.
A linear fit to only the galaxies give a slope of -0.50$\pm$0.04.
The slope obtained with the new sample is  somewhat flatter than 
the slope of -0.65$\pm$0.05 obtained by \citet{odea&baum97}. 

The same dependence calculated using the current model is
also plotted in  Fig.~\ref{fig:nupeakls} for 
jet powers of, $1\times10^{37}$, $1.3\times10^{38}$, $1.3\times10^{39}$,
$1.3\times10^{40}$ and $5\times10^{40}$~W. These curves fit the data well.
The slope of the model curves for $\beta=1.9$, is about -0.85.  
For $\beta=0$ the slope of the model curve is -0.56 which is close to the 
linear fit to the observed data. 
These results suggest that SSA can explain the turnover in the spectrum 
and the observed trend in \nupeak\ --~LS plane can be explained by SSA alone.
The sources evolve from GPS to CSS in the \nupeak\ --~LS plane and do not
leave this plane during the evolution. 
However large CSS sources with low jet power may not show turnover in the observable 
frequency range and may leave this plane during the later stages of evolution.

Using the radio source model of \citet{begelman96} and assumption
on the variation of the magnetic field, B, 
\citet{odea&baum97} predict a slope from -1.65 to -1.8 depending on 
their model parameters. The is steeper than that given by this
model.  Interestingly the slope predicted  by
\citep{bicknell.etal97} for the case of FFA, for $\beta=1.9$
is about -0.91. This is very similar to the 
slope predicted by the current model with $\beta=1.9$.
For $\beta=0.0$, the current model prediction fits the observed trend well.
It is not possible to favour any one mechanism from 
this slope alone.  However, if both mechanisms are at work, 
then the spectral index on the optically thick side is expected to be 
steeper than that predicted by  either of the mechanisms.  
Although in general this is not the case, it is interesting observationally. 
Detailed modelling of the spectrum of high resolution observations 
of a few GPS sources suggest that both mechanisms could be at work
\citep{xie.etal05, kameno.etal03a, kameno.etal03b}.

\subsection{Luminosity evolution}

\begin{figure}
  \includegraphics[width=8.5cm]{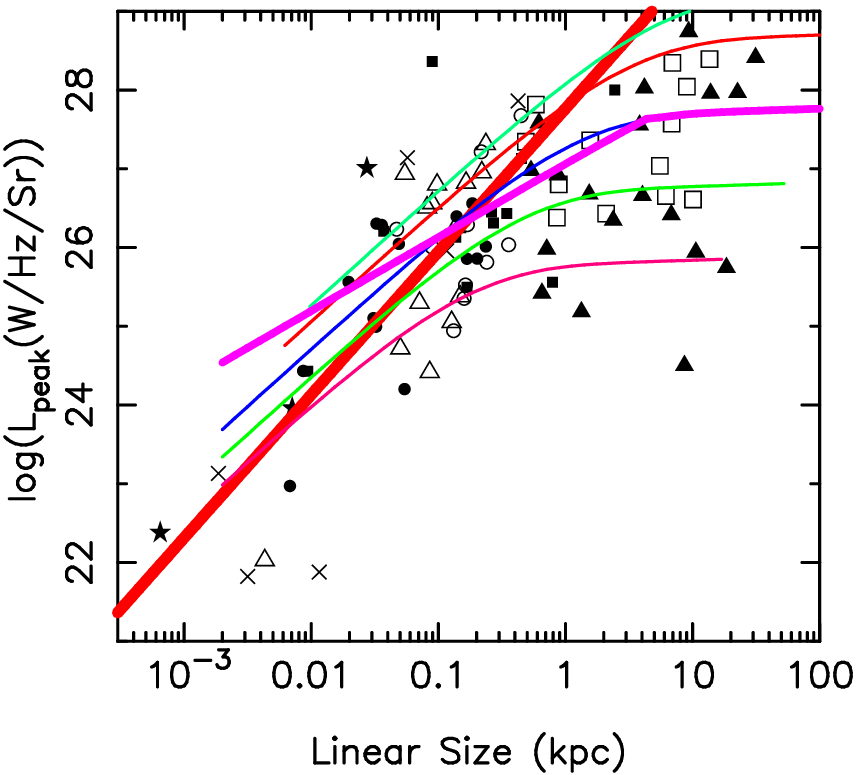}
  \caption{The \lpeak\  is plotted against the LS for the galaxies from
Tables~\ref{table:samplegps} and ~\ref{table:samplecss}.
The meaning of the symbols and the line types are as in Fig.~\ref{fig:nupeakls}.
\label{fig:lsizelum} }
\end{figure}

\begin{figure}
  \includegraphics[width=8.5cm]{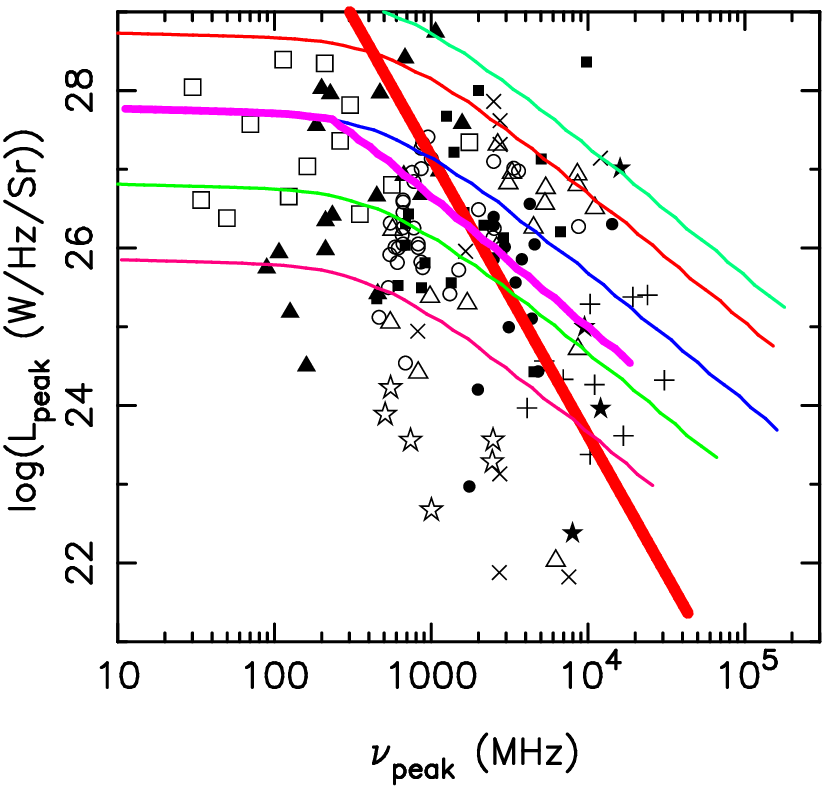}
  \caption{The \lpeak\  is plotted against the \nupeak\  for the galaxies from
Tables~\ref{table:samplecss} and ~\ref{table:samplegps}.
The meaning of the symbols and the line types are as in Fig.~\ref{fig:nupeakls}.
\label{fig:nupeaklum} }
\end{figure}

The vaule of \lpeak\  is plotted against the LS in Fig.~\ref{fig:lsizelum}. 
for sources from Table~\ref{table:samplegps}. 
Since the luminosities of the quasars are affected by the beaming effects, 
only galaxies are plotted in the above diagram. 
The figure shows that the \lpeak\  increases with the LS for the sources with 
sizes less than about 1~kpc. Beyond this size the \lpeak\  does not
vary with LS. A parabolic fit to the data (not shown here) 
clearly shows a flattening at large sizes. A linear least squares fit to 
the data for log(LS) smaller than -0.5  is shown as thick solid line. 
The slope of the fit is 1.8$\pm$0.2. 
The results from the model are also plotted  in the figure for the
same parameters used in the above section. The slope of
the model curves at smaller linear sizes for
$Q_0=1.3\times10^{39}$~W and $\beta$ of 0 and 1.9 are 0.94  and 1.41 
respectively. The slope is higher for higher jet powers. 
This trend is also consistent with  the prediction of \citet{snellen.etal00}.
Although the trend shown by the model curves are in agreement 
with the observed trend, the observations suggest a steeper evolution 
than the model prediction.

It is possible to study this luminosity evolution in the \lpeak\ --~\nupeak\  plane,
where all the GPSs without angular  size estimates can be used. This will
increase the number of sources with very high frequencies or correspondingly
small linear sizes. The dependence of \lpeak\  with \nupeak\  is shown
in Fig.~\ref{fig:nupeaklum} for all the galaxies. 
The figure shows a trend of \lpeak\  decreasing with \nupeak. 
The relations \nupeak\  $\propto$~LS$^{-0.51}$ and \lpeak\  $\propto$~LS$^{1.8}$
obtained from the least squares fits above can be translated to 
\lpeak\  $\propto$~\nupeak$^{-3.5}$.  This trend is shown as a thick solid line.
Such a trend is expected for a power-law electron energy spectrum 
since the $\gamma$ of the electrons  
corresponding to the \nupeak\  is higher and the  electron density is lower
for higher $\gamma$ values.  However the slope expected from this 
argument is the spectral index $\alpha$ which is smaller than the least
squares fit. 
The curves of the model results are plotted for the same 
parameters used for the model
curves in Fig.~\ref{fig:nupeakls}.  
The model curves show that the peak luminosity decreases with 
increasing \nupeak\  for values of \nupeak$~> \nu_{cut}$. The
value of $\nu_{cut}$ is the turnover frequency expected for a source size of
about 1 kpc. For smaller \nupeak\  values corresponding to larger sources the \lpeak\  
is almost constant. 
The slope of the model curves 
for $\beta=1.9$ and 0 is -1.58 and -1.67 respectively. This reflects
the same trend seen in the \lpeak\ --~LS plane.
Most of the sources in this diagram can be bound by curves corresponding 
to jet powers between $1.3\times10^{37}$ and $5\times10^{40}$~W. 
The observed trend and the model curves suggest that
the luminosity increases as the source evolves from GPS to CSS 
and it is unlikely to observationally miss the CSS sources due to evolution.
However it is possible to miss the smaller GPSs since they will appear 
more fainter at the frequencies away from the \nupeak.  
These missing sources do not affect the trend seen in Fig.~\ref{fig:nupeakls}. 

\section{Conclusions}
\label{sec:conclusions}
The dependence of \nupeak\  with  linear size 
is presented for the CSS and GPS sources derived from complete 
samples available in the literature.
The \nupeak\ is anti-correlated with the linear size, 
as \nupeak\ $\propto$~LS$^{-0.51}$ for this sample. 
The sources evolve in luminosity as they grow in linear size. 
The \lpeak\  increases with the LS of the source for sizes smaller
than about 0.3~kpc, as \lpeak\  $\propto$~LS$^{1.8}$. 
Beyond this size the \lpeak\  is almost constant.
This luminosity evolution is seen in the \lpeak\ -- \nupeak\  plane also.
Optical depth effect has been included to the 3D model of K00. 
Using this model the observed dependence of the \nupeak\  with linear size
and \lpeak\  can be explained.  These results suggest that 
synchrotron self absorption can explain the turnover 
in the young radio sources.  The luminosity evolution does not affect
the trend  seen in the \nupeak\ --~LS plane.

\section*{Acknowledgments}

Critical comments and suggestions made by the anonymous referee,
which have helped in refining the interpretation and also improving the presentation are acknowledged.
SJ thanks Paul J. Wiita for useful comments on the paper.
This research has made use of the NASA/IPAC Extragalactic Database (NED)
which is operated by the Jet Propulsion Laboratory, California Institute of 
Technology, under contract with the National Aeronautics and 
Space Administration. The authors made use of the database, CATS 
\citep{verkhodanov.etal97} of the Special Astrophysical Observatory.
This work was partially supported by PROMEP/103-5/07/2462 and Conacyt CB-2009-01/130523 grants.
\bibliographystyle{mn2e}
\bibliography{radio}

\label{lastpage}
\end{document}